\documentclass[12pt]{elsart3-1} 
 
\usepackage{amssymb,amsmath,color}
\usepackage[english,francais]{babel}

\usepackage{epsfig}
\usepackage{epstopdf}
\usepackage{framed}
\usepackage[T1]{fontenc}

\newcommand{\remove}[1]{}
\def\be{\begin{equation}}
\def\ee{\end{equation}}
\def\bea{\begin{eqnarray}}
\def\eea{\end{eqnarray}}
\def\ba{\begin{array}}
\def\ea{\end{array}}

\def\bM{{\bf M}}

\def\bx{{\mathbf {x} }}

\let\<\langle
\let \>\rangle

\newcommand{\br}{\boldsymbol{r}}
\newcommand{\bGam}{\boldsymbol{\Gamma}}
\newcommand{\bom}{\boldsymbol{\omega}}
\newcommand{\bgam}{\boldsymbol{\gamma}}
\newcommand{\bsigma}{\boldsymbol{\Sigma}}
\newcommand{\bpsi}{\boldsymbol{\Psi}}

\newcommand{\bbeta}{\boldsymbol{\beta}}

\newcommand{\bmu}{\boldsymbol{\mu}}
\newcommand{\bGamma}{\boldsymbol{\Gamma}}
\newcommand{\bOmega}{\boldsymbol{\Omega}}
\newcommand{\brho}{\boldsymbol{\rho}}

\newcommand{\bkappa}{\boldsymbol{\kappa}}
\def\bOm{\boldsymbol{\Omega}}

\def\boldeta{\boldsymbol{\eta}}

\newcommand{\bfib}{\color{blue}\bfseries\itshape}
\newcommand{\comment}[1]{\vspace{1 mm}\par 
\marginpar{\large\underline{}}\noindent
\framebox{\begin{minipage}[c]{0.95 \textwidth}
{\bfib #1} \end{minipage}}\vspace{1 mm}\par}
\journal{the Acad\'emie des sciences}

\begin{document}
\centerline{}
\begin{frontmatter}

\title{Nonlocal orientation-dependent dynamics of molecular strands}
\selectlanguage{english}
\author[authorlabel1]{Darryl D. Holm},
\ead{d.holm@imperial.ac.uk}
\author[authorlabel2]{Vakhtang Putkaradze},
\ead{putkarad@math.colostate.edu}

\address[authorlabel1]{Department of Mathematics, Imperial College London, London
SW7 2AZ, UK\\
Computer and Computational Science Division,
Los Alamos National Laboratory, Los Alamos, NM, 87545 USA}
\address[authorlabel2]{Department of Mathematics, Colorado State University,
Fort Collins, CO 80523 USA\\
Department of Mechanical Engineering, University of New Mexico, 
Albuquerque NM 87131 USA
}

\bigskip
\remove{
\begin{center}
{\small Received *****; accepted after revision +++++\\
Presented by £££££}
\end{center}
}

\begin{abstract}
Time-dependent Hamiltonian dynamics is derived for a curve (molecular strand) in $\mathbb{R}^3$ that experiences both nonlocal (for example, electrostatic) and elastic interactions. The dynamical  equations in the symmetry-reduced variables are written on the dual of the semidirect-product Lie algebra $so(3)\,\circledS\,(\mathbb{R}^3\oplus\mathbb{R}^3\oplus\mathbb{R}^3\oplus\mathbb{R}^3)$ with three 2-cocycles. We also demonstrate that the nonlocal interaction produces an interesting new term deriving from the coadjoint action of the Lie group $SO(3)$ on its Lie algebra $so(3)$. The new filament equations are written in conservative form by using the corresponding coadjoint actions. 
\selectlanguage{english}
\\ 
{\it To cite this article: D. D. Holm and V. Putkaradze, 
C. R. Acad. Sci. Paris, Ser. I XXX (2008).}

\vskip 0.5\baselineskip
\selectlanguage{francais}
\noindent{\bf R\'esum\'e} \vskip 0.5\baselineskip \noindent
Nous d\'erivons la dynamique Hamiltonienne d'une courbe (chaine mol\'eculaire) dans l'espace physique $\mathbb{R}^3$ sujette \'a des interactions \'elastiques ainsi que non-locales (\'electrostatiques par exemple). Les \'equations dynamiques des variables r\'eduites par sym\'etrie sont \'ecrites sur l'espace dual de l'alg\`ebre de Lie 
$so(3)\,\circledS\,(\mathbb{R}^3\oplus\mathbb{R}^3\oplus\mathbb{R}^3\oplus\mathbb{R}^3)$ (produit semidirect) avec trois 2-cocycles. Nous d\'emontrons aussi que l'interaction non-locale produit  un nouvel terme int\'eressant, qui d\'erive de l'action coadjointe du group de Lie $SO(3)$ sur son alg\'ebre $so(3)$. Les nouvelles \'equations du filament sont \'ecrites sous une forme conservative gr\^ace aux actions coadjointes correspondantes. 
\\
{\it Pour citer cet article~: D. D. Holm and  V. Putkaradze, 
C. R. Acad. Sci. Paris, Ser. I XXX (2008). }

\end{abstract}
\end{frontmatter}

\vspace{-10mm} 
\selectlanguage{francais}
\section*{Version fran\c{c}aise abr\'eg\'ee}
Cette note d\'epasse le cadre des approches \`a la Kirchoff pour d\'eriver la dynamique Hamiltonienne d'une courbe (chaine mol\'eculaire) dans l'espace physique $\mathbb{R}^3$, lorsqu'elle est sujette \`a des interactions \'elastiques ainsi que non-locales (\'electrostatiques par exemple). Cette note s'inspire d'une extension de la th\'eorie des {\it barres g\'eom\'etriques parfaites} \cite{SiMaKr1988}, qui pr\'esente l'interaction \'elastique dans un contexte g\'eom\'etrique par l'usage des {\it coordonn\'ees mat\'erielles}. Cette th\'eorie a d\'ej\`a \'et\'e utilis\'ee pour la description de quelques aspects de la dynamique des prot\'eines \cite{BiCoZh2004}. Toutefois la g\'en\'eralisation des ces th\'eories aux interactions {\it non-locales} (qui d\'ependent du temps) exige l'application de m\'ethodes g\'eom\'etriques pour d\'eriver le principe d'action (\ref{deltaltotal}), qui vont au-del\`a de l'approche \`a la Kirchoff. Les \'equations dynamiques des variables r\'eduites par sym\'etrie (\ref{sigmavar1},\ref{psivar}) sont \'ecrites dans l'espace dual de l'alg\`ebre de Lie 
$so(3)\,\circledS\,(\mathbb{R}^3\oplus\mathbb{R}^3\oplus\mathbb{R}^3\oplus\mathbb{R}^3)$
 (produit semidirect) avec trois 2-cocycles. De plus, les nouvelles \'equations du filament sont \'ecrites sous une forme conservative (\ref{consgen}) en utilisant les actions coadjointes correspondantes. Notre approche rend possible l'inclusion coh\'erente des effets \'electrostatiques et inertiels dans les \'etudes th\'eoriques et num\'eriques de la dynamique des chaines biologiques. Cela garantit aussi la conservation de l'\'energie et \'etend les traitements Hamiltoniens \`a la Lie Poisson \cite{SiMaKr1988,Gay-BaRa2007,Ho2001} afin d'englober la  d\'ependance non-locale des variables.\selectlanguage{english}

\vspace{-6mm} 
\section{Introduction}
\vspace{-3mm} 
In contemporary science, two basic approaches are taken in describing the dynamics of biological strands (such as proteins). These two approaches might be termed \emph{molecular dynamics} and \emph{geometric dynamics}. Molecular dynamics treats a biological molecule as a collection of charged masses interacting in force fields; the equations of motion are derived from Newton's Second Law, with proper addition of random forces. Using this method, scientists have been able to accurately model the realistic dynamics of proteins with complex shapes \cite{SnNgPaGr2002}. Even though this approach has been successful, it poses tremendous demands on computing power; it also limits the ability to achieve \emph{theoretical understanding}  of the dynamics.  
In addition, the computation of folding of even the
most basic proteins using such detailed dynamics challenges the limits  of most modern computers.  This is due to the necessity of modeling the motion of every single atom in the protein as well as using the atomistic stochastic approach for computing resistance and self-interaction of the molecular strand in water, or other solvent. 

Alternatively, consideration of a protein or biological strand as a continuum curve has a long and rich history with some notable successes \cite{GoPoWe1998,BaMaSc1999,GoGoHuWo2000,HaGo2006}. Most considerations in this approach -- both in the investigation of stationary and time dependent models -- have been made in the framework of the Kirchhoff model of elastic rods. That framework, however, has been so far not been able to overcome the mathematical difficulties of incorporating electrostatic (nonlocal) effects into the dynamics, by a straight-forward extension of Kirchhoff theory. Some recent progress has been made on the \emph{stationary} solutions in the continuum framework \cite{ChGoMa2006}. However,  no time-dependent theory considering nonlocal effects yet exists to our knowledge. The difficulties arise because Kirchhoff's theory is formulated in the \emph{intrinsic} frame connected with the deformed rod. Hence, the computation of distances in Euclidean space becomes non-trivial, thereby increasing the difficulty of constructing a consistent time-dependent theory. 

This paper overcomes the problems confronted in standard Kirchhoff-based approaches by using an extension of the theory of \emph{exact geometric rods} \cite{SiMaKr1988}, which puts elastic 
interactions into a geometric framework using the \emph{material frame} viewpoint.  This theory has already been used to
describe some aspects of protein dynamics \cite{BiCoZh2004}. 
However, the generalization of these theories for \emph{time-dependent} nonlocal interactions requires geometric methods that go beyond the Kirchhoff approach. 
The paper considers only inertial effects. 
The introduction of dissipation into the motion of biological strands 
is a complicated issue \cite{GoPoWe1998,GoGoHuWo2000} that will be deferred to future work. \smallskip

{\bf Acknowledgements.} 
The authors were partially supported by NSF grant NSF-DMS-05377891, the US Department of Energy, Office of Science, Applied Mathematical Research, the Royal Society of London Wolfson Research Award and the MISGAM program of the European Science Foundation.

\vspace{-7mm}
\section{Motion of exact self-interacting geometric rods} 
\begin{figure} [h]

\centering 

\includegraphics[width=.6\textwidth]{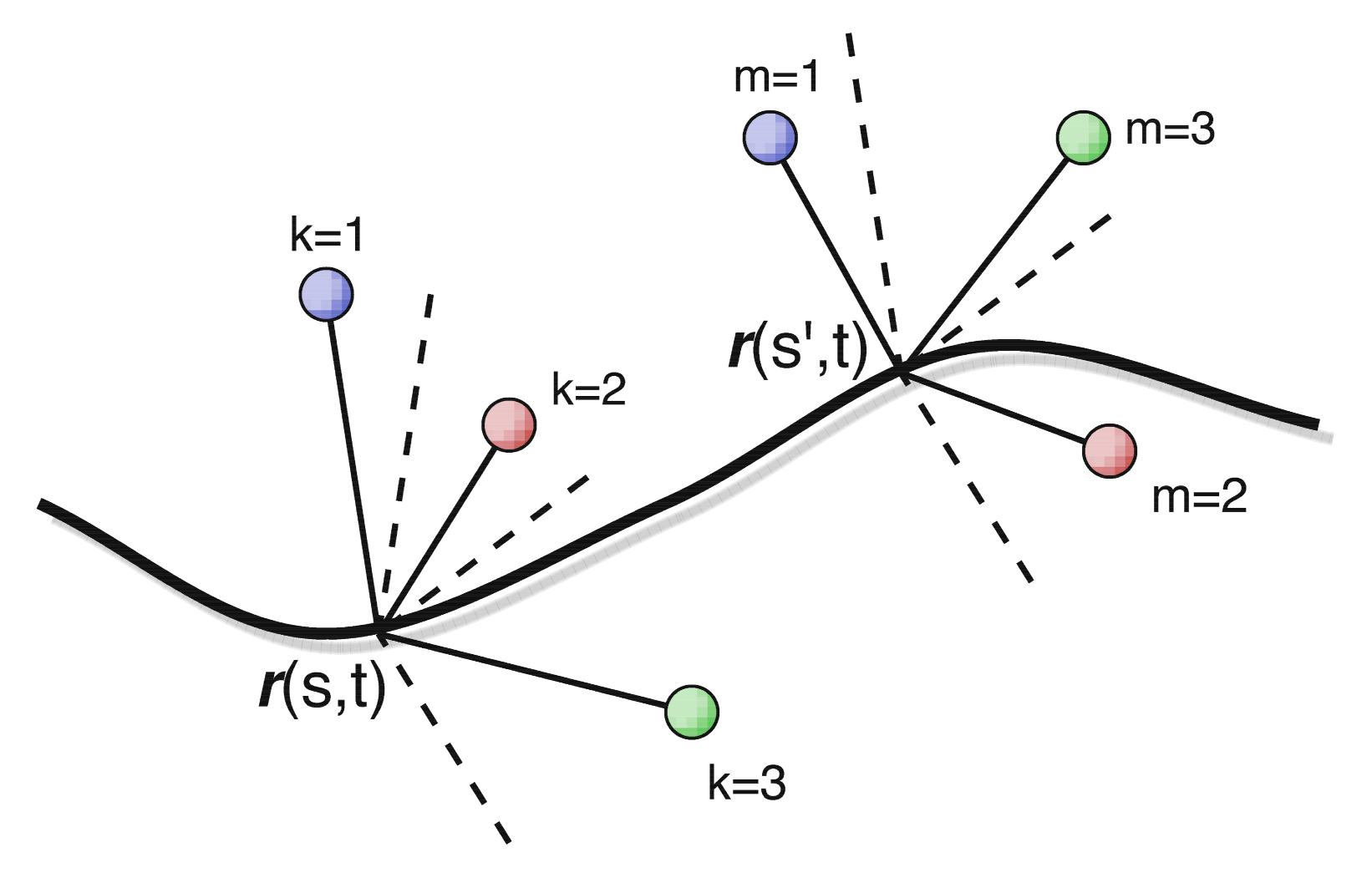} 
 \caption{Rigid conformations of charges are distributed along a curve. } 
 \label{model-fig} 
 \end{figure}

\vspace{-3mm} 
\label{moving-filament}
We consider rigid conformations of charges mounted along a flexible filament at $\br(s,t)$ at distances $\boldeta_k(s,t)$ and allow these charges to interact with each other via a potential (for proteins, the screened electrostatic potential). For simplicity, each charge `bouquet' is assumed to rotate as a rigid body with respect to its origin. 
This rigid conformational rotation is illustrated in Figure \ref{model-fig}.
The dependence on time $t$ of the orientation of the rigid conformation of charges at a spatial point $\br(s,t)$ along  the filament is denoted as  $\Lambda(s,t)\in SO(3)$.
Suppose each rigid conformation of charges is identical and the $k$-th electrical charge is positioned near a given point $\br(s,t)$ of the curve at the reference state $\br+\boldeta_k(s)$. Here, $\boldeta_k(s)$ is a vector of constant length that determines the position of the $k$-th electrical charge relative to the point $\br(s,t)$ along the curve in its reference configuration. If the curve position $\br$ remains fixed, rotation is allowed only in a plane so $\Lambda \in SO(2)$  and there is only one charge $k=1$, our model reduces to that considered in \cite{Mezic2006}. In general,  the position $\mathbf{c}_k$ of the $k$-th charge in the rigid conformation anchored at position $\br(s,t)$ rotates to a new position with $\Lambda(s,t) \in SO(3)$ as
\begin{equation} 
\mathbf{c}_k(s,t)=\br(s,t)+\Lambda(s,t) \boldeta_k(s)
\,,\quad\hbox{where}\quad
\Lambda(s,0) = {\rm Id}
\,.
\label{charge-position}
\end{equation} 
The key to further progress is to use maximum possible reduction of the Lagrangian to the $SO(3)$-invariant quantities. As we show below, because of the nonlocality, the complete reduction from the Lie group to its Lie algebra is impossible, yet the equations can still be formulated as motion on a Lie algebra with an elegant mapping from the unreduced Lie group terms to the Lie algebra. 

The nonlocal part of the potential energy $l_{np}$ of interaction between rigid conformations of charges  at spatial coordinates $s$ and $s'$ along the filament depends only on the distance $d_{k,m} (s, s')$ between the $k$-th and $m$-th charges in the two conformations, 
\begin{equation} 
l_{np}=\sum_{k,m} \frac{1}{2} 
\int U\Big( d_{k,m} (s,s')  \Big) 
\mbox{d} s
\mbox{d} s' 
\, , 
\quad \quad 
d_{k,m}(s,s')=\left| \mathbf{c}_k(s) -\mathbf{c}_m(s') \right| 
\,.
\label{Energy0} 
\end{equation} 
This scalar distance may also be expressed in terms of vectors seen from the frame of orientation of the rigid body at a point $\bx$ along the filament, as
\begin{equation} 
d_{k,m}(s,s')
=\left| \mathbf{c}_k(s) -\mathbf{c}_m(s') \right| =
\remove{ 
\\
&=\left|
 \Lambda^{-1}(s) \left( \mathbf{c}_k(s) -\mathbf{c}_m(s') \right)  
 \right| 
\nonumber 
\\ 
&= 
\left|
 \Lambda^{-1}(s) \left(\br(s)-\br(s')')+ \boldeta_k(s) -\Lambda^{-1}(\bx) \Lambda(s')\boldeta_m(s') \right)  
 \right|  
 \nonumber 
 \\ 
 & =
 }  
 \left| 
\bkappa(s,s')+ \boldeta_k(s) -\xi(s,s') \boldeta_m (s')
 \right|
 \, , 
\label{LiePoissondist} 
\end{equation} 
\begin{equation} 
\mbox{where} \quad 
\bkappa(s,s')
= 
\Lambda^{-1}(s) \big(\br(s) -\br(s') \big) 
\in \mathbb{R}^3
\quad\hbox{and}\quad
\xi(s,s')=  \Lambda^{-1}(s) \Lambda(s') \in SO(3)
\,.
\label{rho-xi-defs}
\end{equation}
The first of these quantities is the spatial vector $\br(s)-\br(s')$ between two points on the filament, as seen from the orientation $\Lambda(s)$ of the rigid body (charge conformation) at $s$ on the filament. The second is the \emph{relative} orientation of the rigid bodies (charge conformations) at $s$ and $s'$. A transposition identity,
$\xi(s,s')^T=\xi(s',s)=\xi(s,s')^{-1},
$
follows from the definition of $\xi(s,s')$ in (\ref{rho-xi-defs}).

\begin{rem} 
Both the vector $\bkappa(s,s')$ and the relative orientation $\xi(s,s')$ defined in (\ref{rho-xi-defs}) are invariant under changes of the orientation of the spatial coordinate system obtained by the left action $\br(s)-\br(s')\to O(\br(s)-\br(s'))$ and $\Lambda\to O\Lambda$ of any element $O$ in the rotation group $SO(3)$. 
\end{rem} 

The localized part of the Lagrangian depends on $\Lambda$, ${\dot\Lambda}$, $\Lambda'$, $\br(s)$, $\dot{\br}$, $\br'$, where the dot denotes the time derivative and prime is the derivative with respect to arclength $s$. If it is left-invariant under the action of $SO(3)$, this Lagrangian may be reduced to a function of left-invariant quantities:
$\Omega=\Lambda^{-1} \Lambda'  \in so(3)$,  
$\omega=\Lambda^{-1} \dot{\Lambda} \in so(3)$, 
$\bGam=\Lambda^{-1} \br'   \in \mathbb{R}^3$, 
$\bgam=\Lambda^{-1} \dot{\br}  \in \mathbb{R}^3$ and 
$\brho=\Lambda^{-1} \br   \in \mathbb{R}^3$. 
Capital Greek letters denote derivatives in $s$, while lower-case Greek letters (except for $\brho$) denote derivatives in time. Bold letters such as $\bom$ denote vectors in $\mathbb{R}^3$ while $\Omega$ is a skew $3\times3$ matrix in the Lie algebra $so(3)$ whose entries correspond to vector components via the isomorphism between $so(3)$ and $\mathbb{R}^3$. For example,  for any vector $\boldsymbol{v}\in\mathbb{R}^3$ one has $\Omega \boldsymbol{v}=\bOm \times \boldsymbol{v}$. 
In components, this is the map ${\Omega}_{jk}
=
-\,\epsilon_{jkl}\Omega_l$ without any extra adornments.

In terms of these quantities, we assume the symmetry-reduced Lagrangian $L$ may be written as the sum of a local  part $l$ and a nonlocal part $l_{np}$, according to 
\begin{equation} 
L(\brho,\bgam,\bGam,\bom,\bOm,\xi)= 
l(\brho,\bgam,\bGam,\bom,\bOm)
+ \int 
U \big( 
\bkappa(s,s'), \xi(s,s')
\big) 
\mbox{d} s 
\mbox{d} s' 
:= 
l+l_{np}
\, ,
\label{totalL}
\end{equation} 
where the left-invariant quantity $\bkappa(s,s')$ in (\ref{rho-xi-defs}) may be expressed using $\brho$ and $\xi$ as
\begin{equation} 
\bkappa(s,s') := \Lambda^{-1} (s)\big( \br(s) - \br(s') \big)
= \brho(s) - \xi(s,s') \brho(s') 
\label{kappadef} 
\, .
\end{equation}

{\bf Kinematics}. 
Let us compute the space and time derivatives of $\brho=\Lambda^{-1} \br   \in \mathbb{R}^3$. The space ($s$) derivative of $\brho$ (denoted by a prime) and time ($t$) derivative (denoted by a dot) are given  by
\begin{equation} 
\brho' = -\, \Omega \brho +\bGam
=-\,\bOm \times  \brho +\bGam
\quad \mbox{and} \quad 
\dot{\brho} = -\, \omega \brho +\bgam
=-\, \bom \times  \brho +\bgam
\,.
\label{rhotimederiv} 
\end{equation}
Compatibility of these formulas arises from equality of the cross-derivatives of $\br$ and $\Lambda$. Namely, 
 \remove{
Namely, 
\[ 
\bgam\, '=- \bOm \times \bgam + \Lambda^{-1} \dot{\br ' } 
\,,
\] 
and 
\[ 
\dot{\bGam} '=- \bom \times \bGam + \Lambda^{-1} \dot{\br ' } 
\,.
\] 
By subtracting these equations, we obtain the following kinematic relation 
}
\begin{equation} 
\dot{\bGam}+ \bom \times \bGam = 
\bgam\,'+ \bOm \times \bgam 
\quad\hbox{and}\quad
\dot{\bOm} =
 \bOm  \times \bom + \bom\,'
\, . 
\label{kincond}  
\end{equation} 

{\bf Variations.} 
The variations of $\brho$, $\bom$, $\bgam$, $\bOm$ and $\bGam$ are computed by the following steps: 
\begin{equation} 
\delta \brho =  
- \Lambda^{-1} \delta \Lambda \Lambda^{-1} \br
+ \Lambda^{-1} \delta \br 
= 
- \Sigma \brho +\bpsi 
=  
- \bsigma \times \brho + \bpsi 
=  \brho \times \bsigma  + \bpsi 
\,,
\label{rhovar} 
\end{equation} 
where one defines the left-invariant variations
$\Sigma= \Lambda^{-1} \delta \Lambda   \in so(3)$ and 
$\bpsi=\Lambda^{-1} \delta \br \in \mathbb{R}^3$. 
The variations $\delta \bgam$, $\delta \bGam$, $\delta \bom$ and  $\delta \bOmega$ are given in terms of the the left-invariant quantities $\Sigma$ and $\bpsi$ by 
\begin{equation} 
\delta \bgam =
 -\,\bsigma \times \bgam 
 + \bom \times \bpsi 
 +\frac{\partial \bpsi}{\partial t}
  =
 \bgam \times \bsigma 
 + \bom \times \bpsi 
 +\frac{\partial \bpsi}{\partial t} 
 \, , 
 \label{gammavar} 
\end{equation} 
\remove{
\[ 
\delta \bGam =
  - \Lambda^{-1} \delta \Lambda \Lambda^{-1} \brho' + 
\Lambda^{-1} \delta \rho' 
=
 -\Sigma \bGam + \Omega \bpsi +\frac{\partial \bpsi}{\partial s}  \, , 
\] 
}
\begin{equation} 
\delta \bGam =
 -\,\bsigma \times \bGam 
 + \bOm \times \bpsi 
 +\frac{\partial \bpsi}{\partial s} 
 =
 \bGam \times \bsigma 
 + \bOm \times \bpsi 
 +\frac{\partial \bpsi}{\partial s} 
 \, , 
 \label{Gammavar} 
\end{equation} 
\remove{ 
Next, 
\[ 
\delta \omega =
  - \Lambda^{-1} \delta \Lambda \Lambda^{-1} \dot{\Lambda} + 
\Lambda^{-1} \delta \dot{\Lambda}
 = 
 -\Sigma \omega + \omega \Sigma +\frac{\partial \Sigma}{\partial t} = 
 [\omega, \Sigma]+ \frac{\partial \Sigma}{\partial t} 
 \, , 
\] 
so expressing these formulas in terms of vectors yields
}

\remove{
\[ 
\delta \Omega =
  - \Lambda^{-1} \delta \Lambda \Lambda^{-1} \Lambda ' + 
\Lambda^{-1} \delta \Lambda' 
= 
 -\Sigma \Omega + \Omega \Sigma +\frac{\partial \Sigma}{\partial s} = 
 [\Omega, \Sigma]+ \frac{\partial \Sigma}{\partial s} \, , 
\] 
so, again, expressing in terms of vectors leads to
}
\begin{equation} 
\delta \bom =
 \bom  \times \bsigma 
 +
 \frac{\partial \bsigma}{\partial t}  
 \quad \hbox{and} \quad 
\delta \bOm =
 \bOm  \times \bsigma +\frac{\partial \bsigma}{\partial s} 
 \, . 
 \label{omegavar} 
\end{equation} 
The key to understanding the nonlocal variations lies in the matrix formula
\begin{equation} 
\xi^{-1} \delta \xi(s,s') =-{\rm Ad}_{\xi^{-1}(s,s')} \Sigma (s) + \Sigma(s') \, , \quad\hbox{where}\quad
{\rm Ad}_{\xi^{-1}} \Sigma := \xi^{-1} \Sigma\, \xi
\,,
\label{xivar3} 
\end{equation} 
obtained from the definition of $\xi(s,s')$ in equation (\ref{rho-xi-defs}).
The variation of $\bkappa$ in (\ref{kappadef}) is then given by 
\begin{equation} 
\delta \bkappa(s,s') = 
\remove{ 
& - \Sigma(s)  \bkappa(s,s') 
+ \bpsi(s) 
- \xi(s,s') \bpsi(s') 
\nonumber 
\\ 
= 
}
 -\, \bsigma(s)  \times \bkappa(s,s') 
+ \bpsi(s)-\xi(s,s') \bpsi(s')
\, . 
\label{deltakappa} 
\end{equation}

\remove{
Suppose now we want to compute variations of reduced Lagrangian energy $l$ that is a functional of 
$(\brho, \bgam, \bGam, \omega, \Omega)$. From (\ref{LiePoissondist}) we see that 

\emph{Note.} From now on, we assume that the nonlocal part of the potential energy $U$  is a function of the two variables $\bkappa(s,s')$ and $\xi(s,s')$. In particular, for a potential energy depending on the distance $d_{k,m}$, the variables  $\bkappa$ and $\xi$ enter in the linear combination defined by $d_{k,m}$. In principle, the potential energy could have chosen to be an arbitrary functional of $\brho(s)$  and $\xi(s,s')$. Euler-Poincar\'e methods would be directly applicable to these functionals as well. 
\smallskip 
}

{\bf Calculation of energy variations.}
The equations of motion are computed from the stationary action principle $\delta S=0$, with $S=\int L\,dt$ and $L=l+l_{np}$ in equation (\ref{totalL}), for which 
\begin{align} 
\delta S= & \int 
\left< 
\frac{\delta l}{\delta \brho}
\, , \, 
\delta \brho 
\right> 
+ 
\left< 
\frac{\delta l}{\delta \bgam}
\, , \, 
\delta \bgam 
\right> 
+
\left< 
\frac{\delta l}{\delta \bGam}
\, , \, 
\delta \bGam
\right> 
+ 
\left< 
\frac{\delta l}{\delta \omega}
\, , \, 
\delta \omega
\right> 
\nonumber 
\\
&
+
\left< 
\frac{\delta l}{\delta \Omega}
\, , \, 
\delta \Omega
\right> 
+
\left< 
\frac{\delta l_{np}}{\delta \bkappa}
\, , 
\delta \bkappa
\right> 
+
\left< 
\xi^{-1} \frac{\delta l_{np}}{\delta \xi}
\, , 
\xi^{-1} \delta \xi
\right> \mbox{d} t =0 \, . 
\label{deltaltotal} 
\end{align} 
The terms proportional to $\bsigma$ and $\bpsi$ give, respectively, 
\begin{align}
\bigg(\frac{\partial}{\partial t} \frac{\delta l}{\delta \bom}
\
+\
\bom \times \frac{\delta l}{\delta \bom}
\bigg)
+
\left(
\frac{\partial}{\partial s} \frac{\delta l}{\delta \bOm}
+
\bOm \times \frac{\delta l}{\delta \bOm}
\right) 
&= \frac{\delta l}{\delta \bgam} \times \bgam
+
\frac{\delta l}{\delta \bGam} \times \bGam
+
\frac{\delta l}{\delta \brho} \times \brho 
\label{sigmavar1} \\
&
+\int \frac{\partial U}{\partial \bkappa} (s,s') \times \bkappa (s,s') \mbox{d} s' 
+
\int \mathbf{Z}(s,s') \mbox{d} s'
\,,
\nonumber
\end{align}
\begin{align}
\bigg(\frac{\partial}{\partial t} \frac{\delta l}{\delta \bgam}
\
+\
\bom \times \frac{\delta l}{\delta \bgam}
\bigg)
+
&
\left(
\frac{\partial}{\partial s} \frac{\delta l}{\delta \bGam}
+
\bOm \times \frac{\delta l}{\delta \bGam}
\right)
=
\nonumber 
\\
& \left(
\frac{\delta l}{\delta \brho}
+
\int \frac{\partial U}{\partial \bkappa} (s,s')
- 
\xi(s,s') \frac{\partial U}{\partial \bkappa} (s',s)
\mbox{d} s'
\right)
.
\label{psivar}
\end{align}

Here, we have defined the nonlocal contribution 
\begin{equation} 
Z(s,s'):=\mathbf{Z}(s,s')\times = \xi(s,s') 
\Big(
\frac{\partial U}{\partial \xi} (s,s') 
\Big)^T
-\frac{\partial U}{\partial \xi} (s,s') 
\xi^T(s,s') 
\, . 
\label{Zdef2}
\end{equation} 
The term $Z(s,s')$ is the contribution from the nonlocal part of the Lagrangian we have sought. A direct calculation shows that $Z^T=-Z$, so $Z \in so(3)$. This expression appears naturally in geometric mechanics of interacting oriented bodies like asteroids \cite{CeMa2005},  and more general theory of \emph{reduction by stages} 
\cite{CeMaRa2001} illuminates the geometric structure  of such interactions. 


\vspace{-6mm} 
\section{Spatial conservation laws}
\vspace{-3mm} 
The physical meaning of equations (\ref{sigmavar1}, \ref{psivar}) is revealed by writing them as spatial conservation laws. 
For this, we invoke the following identities valid for the Ad- and Ad$^*$-actions of any element $g(t)\in G$  in the Lie group on $\eta \in \mathfrak{g}$ in the Lie algebra and $\mu \in \mathfrak{g}^*$ in its dual with pairing $\langle \,\cdot\,, \,\cdot\, \rangle: \mathfrak{g}^*\times\mathfrak{g}\to \mathbb{R}$ 
\begin{equation} 
\Big<\mu,\,{\rm Ad}_{g^{-1}(t)} \frac{\partial }{\partial t} {\rm Ad}_{g(t)} \eta \Big>
=\Big<\mu,\,{\rm ad}_{\sigma}  \eta \Big>
\quad\hbox{and}\quad
\Big< 
{\rm Ad}^*_{g(t)} \frac{\partial}{\partial t} 
{\rm Ad}^*_{g^{-1}(t)}\mu  
\, , \, 
\eta
\Big> 
=
 \Big<
 - {\rm ad}^*_ \sigma \mu 
 \, , \, 
 \eta 
 \Big> 
\,,
\label{diffAd}  
\end{equation} 
where ${\rm Ad}^*: G\times\mathfrak{g}^*\to\mathfrak{g}^*$ is defined by $\langle {\rm Ad}^*_{g(t)}\mu\,,\,\eta\rangle:=\langle \mu\,,\,{\rm Ad}_{g(t)}\eta\rangle$ and $\sigma = g^{-1} g'(t) \in \mathfrak{g}$ belongs to the Lie algebra $\mathfrak{g}$. 
Equations (\ref{sigmavar1}, \ref{psivar}) are formulated on the \emph{dual} of the Lie algebra, for which the second equation in (\ref{diffAd}) gives 
\remove{
Thus, we choose an arbitrary $\psi \in \mathfrak{g}$, $g(t) \in G$ and $\eta(t) \mathfrak{g}$ and proceed and follows. 
}
\begin{equation} 
\Big< 
{\rm Ad}^*_{g(t)} \frac{\partial}{\partial t} 
\big(  {\rm Ad}^*_{g^{-1}(t)}\mu(t) \big) 
\, , \, 
\eta
\Big> 
 = 
  \Big<
 \dot{\mu} - {\rm ad}^*_ \sigma \mu 
 \, , \, 
 \eta 
 \Big> 
 \, . 
 \label{diffAdstar} 
\end{equation} 

To derive the conservation form of equations  (\ref{sigmavar1}, \ref{psivar}), we consider the group $G=SE(3)$ with the group element $g=(\Lambda(s,t) , \br(s,t) )$, whose left-invariant time-derivative is 
$\sigma = ( \Lambda^{-1} \dot{\Lambda}, \Lambda^{-1} \dot {\br}) = (\bom, \bgam).$
Using the definition of ${\rm ad}^*$ for the $se(3)$ Lie algebra yields 
\begin{equation} 
{\rm Ad}^*_{g(t)} 
\frac{\partial}{\partial t} \Big[ 
 {\rm Ad}^*_{g^{-1}(t)} 
 \Big( 
 \frac{\delta l}{\delta \bom} \, , \,  \frac{\delta l}{\delta \bgam} 
 \Big) 
  \Big]
  =
\frac{\partial}{\partial t}
\Big( 
\frac{\delta l}{\delta \bom} \, , \,  \frac{\delta l}{\delta \bgam} 
\Big) 
+ 
\Big( 
\bom \times \frac{\delta l}{\delta \bom} + 
\bgam \times  \frac{\delta l}{\delta \bgam}
\, , \, 
-\, \bom \times  \frac{\delta l}{\delta \bgam}
\Big) 
\, . 
\label{timeAd}
\end{equation}  
The tangent vector obtained from the derivative in arclength $s$ is 
$( \Lambda^{-1} \Lambda', \Lambda^{-1} \br') = (\bOm, \bGam).$ 
Thus, 
\begin{equation} 
{\rm Ad}^*_{g(s)} 
\frac{\partial}{\partial s} \Big[ 
 {\rm Ad}^*_{g^{-1}(s)} 
 \Big( 
 \frac{\delta l}{\delta \bOm} \, , \,  \frac{\delta l}{\delta \bGam} 
 \Big) 
  \Big]
  =
\frac{\partial}{\partial s}
\Big( 
\frac{\delta l}{\delta \bOm} \, , \,  \frac{\delta l}{\delta \bGam} 
\Big) 
+ 
\Big( 
\bOm \times \frac{\delta l}{\delta \bOm} + 
\bGam \times  \frac{\delta l}{\delta \bGam}
\, , \, 
-\, \bOm \times  \frac{\delta l}{\delta \bGam}
\Big) 
\, . 
\label{spaceAd}
\end{equation}  
\begin{rem}
The nonlocal term (\ref{Zdef2}) arises as the derivative of the nonlocal part of the potential with respect to Lie algebra elements $\bOm$ and $\bGam$, as follows. 
\end{rem}
Upon identifying coefficients of the free variations  $\bsigma\times=\Lambda^{-1} \delta \Lambda$ and $\bpsi= \Lambda^{-1} \delta\br$, one may write the following identity relating different variational derivatives of the nonlocal potential $l_{np}$: 
 \[ 
\delta l_{np} = 
 \Big< 
\xi^{-1} \frac{\delta l_{np} }{\delta \xi} 
\, , \, 
\xi^{-1} \delta \xi
\Big> 
+ 
\Big< 
\frac{\delta l_{np} }{\delta \bkappa} 
\, , \, 
\delta \bkappa 
\Big> 
= 
\Big< 
\frac{\delta l_{np} }{\delta \bGam} 
\, , \, 
\delta \bGam 
\Big> 
+ 
\Big< 
\frac{\delta l_{np} }{\delta \bOm} 
\, , \, 
\delta \bOm 
\Big> 
\, . 
\] 
Using expressions (\ref{xivar3}) for $\xi^{-1} \delta \xi$, 
(\ref{deltakappa}) for $\delta \bkappa$, (\ref{omegavar}) for $\delta \bOm$ and 
(\ref{gammavar}) for $\delta \bGam$,  then collecting terms proportional to the free variation $\bsigma$ 
yields the following identity, which implicitly defines $\delta l_{np} / \delta \bOm$  in terms of known quantities, 
\begin{equation} 
-\frac{\partial}{\partial s} \frac{\delta  l_{np}}{\delta \bOm} 
 + \bOm \times \frac{\delta  l_{np}}{\delta \bOm} 
 = 
 \int \frac{\partial U}{\partial \bkappa} (s,s') \times \bkappa (s,s') 
 \,\mbox{d} s' 
+
\int \mathbf{Z}(s,s') \,\mbox{d} s' 
\,,
\label{dlnpdOm} 
\end{equation} 
where we have defined $\mathbf{Z}(s,s')$ according to (\ref{Zdef2}). Likewise, identifying terms multiplying $\bpsi$ gives 
\begin{equation} 
-\frac{\partial}{\partial s} \frac{\delta  l_{np}}{\delta \bGam} 
 + \bOm \times \frac{\delta  l_{np}}{\delta \bGam} = 
 \int \frac{\partial U}{\partial \bkappa} (s,s')
- 
\xi(s,s') \frac{\partial U}{\partial \bkappa} (s',s)
\,\mbox{d} s'
\,.
\label{dlnpdGam} 
\end{equation} 
Therefore, we conclude that equations (\ref{sigmavar1}, \ref{psivar}) are equivalent to the following pair of conservation laws expressed using 
variations of the total Lagrangian $L=l+l_{np}$:  
\remove{ 
\begin{equation} 
\frac{\partial}{\partial t} 
\Big[ 
 {\rm Ad}^*_{g^{-1}(t)} 
 \big( 
 \frac{\delta l}{\delta \bom} \, , \,  \frac{\delta l}{\delta \bgam} 
 \big) 
  \Big]
  + 
\frac{\partial}{\partial s} \Big[ 
 {\rm Ad}^*_{g^{-1}(t)} 
 \big( 
 \frac{\delta l +l_{np}}{\delta \bOm} \, , \,  \frac{\delta l+l_{np}}{\delta \bGam} 
 \big) 
  \Big] =  
  {\rm Ad}^*_{g^{-1}(t)} 
 \Big( 
  \frac{\delta l}{\delta \brho} \times \brho 
  \, , \, 
  \frac{\delta l}{\delta \brho}
  \Big) 
\label{cons} 
\end{equation}  
The terms $\delta l/ \delta \brho$ descirbe the influence of the external forces and torques and thus do not satisfy conservation laws, which is physically sensible. 
Note that $l_{np}$ only depends on $\bOm$ and $\bGam$ by our assumptions. In general, even when both the local and nonlocal potential are present,  the universal conservation law is valid 
}
\begin{equation} 
\frac{\partial}{\partial t} 
\Big[ 
 {\rm Ad}^*_{g^{-1}(t)} 
 \Big( 
 \frac{\delta L}{\delta \bom} \, , \,  \frac{\delta L}{\delta \bgam} 
 \Big) 
  \Big]
  + 
\frac{\partial}{\partial s} \Big[ 
 {\rm Ad}^*_{g^{-1}(t)} 
 \Big( 
 \frac{\delta L}{\delta \bOm} \, , \,  \frac{\delta L}{\delta \bGam} 
 \Big) 
  \Big] =  
  {\rm Ad}^*_{g^{-1}(t)} 
 \Big( 
  \frac{\delta L}{\delta \brho} \times \brho 
  \, , \, 
  \frac{\delta L}{\delta \brho}
  \Big) 
  \,.
\label{consgen} 
\end{equation} 
The terms $(\delta L/ \delta \brho \times \brho,\,\delta L/ \delta \brho)$ describe the influence of the external torques and forces, respectively,   which are not expressible in conservation form.  (This is similar to the situation for the heavy top.)

\begin{rem} 
On Legendre transforming the total Lagrangian $L$ to the Hamiltonian,
\begin{equation}
H(\bmu,\bbeta;\bOmega,\bGamma,\brho)
=
\int (\bmu\cdot \bom + \bbeta\cdot\bgam)\,ds
-
L(\bom,\bgam;\bOmega,\bGamma,\brho)
\,,
\label{Leg-transf}
\end{equation}
equations (\ref{rhotimederiv}), (\ref{kincond}), (\ref{sigmavar1}) and ( \ref{psivar}) may be expressed in Lie-Poisson form as
\begin{equation} 
\frac{\partial}{\partial t}
\left[
\begin{array}{c}
    \bmu
    \\
    \brho
    \\  
    \bOmega
    \\
    \bGamma
    \\
    \bbeta 
    \end{array}
\right] 
= 
\left[
\begin{array}{ccccc}
   \bmu\times
   &\
   \brho\times\
   &
   (\partial_s + \bOmega\times)
   &
   \bGamma\times
   &
   \bbeta\times  
   \\
   \brho\times
   &
   0
   &
   0
   &
   0
   &
   Id
   \\
   (\partial_s + \bOmega\times)
   &
   0
   &
   0
   &
   0
   &
   0
   \\
   \bGamma\times
   &
   0
   &
   0
   &
   0
   &
   (\partial_s + \bOmega\times)
   \\
   \bbeta\times
   &
   -Id\
   &
   0
   &
   (\partial_s + \bOmega\times)
   &
   0
    \end{array}
\right]    
\left[
\begin{array}{c}
   \delta H/\delta\bmu=\bom \\
    \delta H/\delta\brho \\
    \delta H/\delta \bOmega \\
    \delta H/\delta\bGamma  \\
   \delta H/\delta\bbeta=\bgam
    \end{array}
\right] .
\label{LP-Ham-struct-vec-se3}
\end{equation}
This Lie-Poisson Hamiltonian matrix is dual to the semidirect-product Lie algebra $so(3)\,\circledS\,(\mathbb{R}^3\oplus\mathbb{R}^3\oplus\mathbb{R}^3\oplus\mathbb{R}^3)$ with three different types of 2-cocycles defined on its normal $\mathbb{R}^3$ subalgebras. The symplectic 2-cocycle in $\{\brho,\,\bbeta\}$ induces the generalized 2-cocycle in $\{\bGamma,\,\bbeta\}$, for which $\bOmega$ is a Casimir; so $\bOmega\times$ is a constant in this 2-cocycle. 
In contrast, the quantity $\bOmega \times$ in the generalized 2-cocycle for $\{\bmu,\,\bOmega\}$ is a connection form. The latter 2-cocycle also appears in the theory of complex fluids \cite{SiMaKr1988,Gay-BaRa2007,Ho2001}. 
\end{rem}

\begin{rem}
In this paper, the influence of nonlocality (e.g., charge screening) on rod mechanics was studied using the Euler-Poincar\'e variational method.  This variational approach led to an equivalent Lie-Poisson Hamiltonian formulation of the new equations (\ref{sigmavar1}, \ref{psivar}). 
Applying the Ad$^*_{g^{-1}(t)}$ transformation from body to spatial variables in these equations produced a great economy of form and exposed the meaning of the interplay among their various local and nonlocal terms.
\end{rem}

\remove{
\section*{Acknowledgements} 
The authors were partially supported by NSF grant NSF-DMS-05377891. The work of 
DDH was also partially supported  by the US Department of Energy, Office of Science, Applied Mathematical Research, and the Royal Society of London Wolfson Research Merit Award. VP acknowledges the support of  the European Science Foundation for partial support through the MISGAM program. 
}
 
\bibliographystyle{abbrv}
\vspace{-5mm}
\bibliography{HoPu-CR-12Mar2008}

\end{document}